\documentclass[12pt]{article}
\usepackage{a4}
\usepackage{epsf}

\makeatletter
\ifx\undefined\operator@font
  \let\operator@font=\rm
\fi
\def\Re{\mathop{\operator@font Re}\nolimits}
\def\Im{\mathop{\operator@font Im}\nolimits}
\makeatother

\begin{document}

\newcommand{\bq}{\begin{equation}}
\newcommand{\eq}{\end{equation}}
\newcommand{\bqa}{\begin{eqnarray}}
\newcommand{\eqa}{\end{eqnarray}}
\newcommand{\nl}{\nonumber \\}
\newcommand{\mc}{Monte Carlo}
\newcommand{\qmc}{Quasi-Monte Carlo}
\newcommand{\qran}{quasi-random}
\newcommand{\intl}{\int\limits}
\newcommand{\intk}{\intl_K}
\newcommand{\suml}{\sum\limits}
\newcommand{\prol}{\prod\limits}
\newcommand{\umu}{^{\mu}}
\newcommand{\order}[1]{{\mathcal O}\left(#1\right)}
\newcommand{\eqn}[1]{Eq.\ref{#1}}
\newcommand{\intinf}{\intl_{-\infty}^{\infty}}
\newcommand{\intinfi}{\intl_{-i\infty}^{i\infty}}
\newcommand{\ddf}{{\mathcal D}\!f}
\newcommand{\si}{\sigma}
\newcommand{\vn}{\vec{n}}
\newcommand{\svn}{_{\vn}}
\newcommand{\fal}[1]{^{\underline{#1}}}
\newcommand{\bpic}{\begin{picture}}
\newcommand{\epic}{\end{picture}}
\newcommand{\Cross}[4]{\Line(#1,#2)(#3,#4)\Line(#1,#4)(#3,#2)}
\newcommand{\DL}[2]{\DashLine(#1)(#2){1.5}}
\newcommand{\DC}[2]{\DashCArc(#1)(#2,0,360){2}}
\newcommand{\BC}{\BCirc}
\newcommand{\Yv}[1]{\BCirc(#1){3}}
\newcommand{\Vix}[1]{\Vertex(#1){2}}
\newcommand{\avg}[1]{\left\langle #1\right\rangle}

\pagestyle{empty}

\begin{flushright}
NIKHEF 96-03
\end{flushright}

\vspace{2cm}

\begin{center}
\begin{Large}
{\bf Discrepancy-based error estimates\\
 for \qmc.\\
\vspace{\baselineskip}
II: Results in one dimension}\end{Large}\\
\vspace{\baselineskip}
{\bf Jiri Hoogland
     \footnote{e-mail: t96@nikhefh.nikhef.nl,\quad
                       research supported by Stichting FOM}
     and Ronald Kleiss
     \footnote{e-mail: t30@nikhefh.nikhef.nl,\quad
                       research supported by Stichting FOM}
     \\
NIKHEF-H, Amsterdam, The Netherlands}\\
\vspace{2\baselineskip}
{\bf Abstract}
\end{center}
The choice of a point set, to be used in numerical integration,
determines, to a large extent, the error estimate of the integral.
Point sets can be characterized by their discrepancy, 
which is a measure of its non-uniformity.
Point sets with a discrepancy that is low with respect to the expected
value for truly random point sets, are generally thought to be desirable.
A low value of the discrepancy implies a negative correlation
between the points, which may be usefully employed to improve the
error estimate of a numerical integral based on the point set.
We apply the formalism developed in a previous publication to
compute this correlation for one-dimensional point sets, using a
few different definitions of discrepancy.


\newpage
\pagestyle{plain}
\setcounter{page}{1}

\section{Introduction}
In a previous publication \cite{first}, we have discussed the
problem of computing a numerical integral estimate and its error,
using a given set consisting of $N$ points
$x_k$, $k=1,2,\ldots,N$ in the multi-dimensional hypercube $K$.
As is well known, in the \mc\ approach, where the points
are chosen iid random with the uniform distribution, the expected error
decreases as $1/\surd{N}$. This can be improved upon by applying
{\em \qran} point sets, that are, owing to some more or less sophisticated
generation algorithm, more uniformly distributed than random ones.
This is embodied in the notion of {\em discrepancy\/} $D_N$, a
measure of the deviation from uniformity. A very uniformly distributed
point set will have a relatively low value $s$ of this discrepancy,
whereas very non-uniform ones will have a high value of $s$. Obviously,
different definitions of discrepancy are possible in addition to the
best-known so-called {\em star-discrepancy} $D_N^{\ast}$.
Good uniformity implies, however, that the points are not independent,
and in particular the distributions $P_1(x_1)$ of a single point $x_1$, 
and $P_2(x_1,x_2)$, of a pair of points $x_1,x_2$, 
will deviate from unity\footnote{These are the distributions
obtained by integrating over the remaining points.}: we can define
\bqa
P_1(x_1) & = & 1 - {1\over N}F_1(s;x_1)\quad,\nl
P_2(x_1,x_2) & = & 1 - {1\over N}F_2(s;x_1,x_2)\quad.
\eqa
As discussed in \cite{first}, in order to talk sensibly about such
distributions, we must have an underlying definition of an ensemble
of point sets. In the truly random, \mc, case this is the ensemble of
all sets of $N$ points with the Cartesian product of $N$ uniform
one-point distributions: in the \qran\ case, we restrict this ensemble
to all point sets having a fixed value $s$ of the discrepancy. For
truly random points, $s$ is then a random variable, with a probability
density $H_0(s)$. In \cite{first}, we have developed a formalism that
allows us to define a useful form for $D_N$, and from that to compute
a $1/N$ expansion for $H_0(s)$, $F_1(s;x_1)$ and $F_2(s;x_1,x_2)$.
Let $f(x)$ be a quadratically integrable function on $K$.
For a general integration error
\bq
\eta = {1\over N}\suml_{k=1}^Nf(x_k) - \intk dx\;f(x)
\eq
we can then show that, averaged over the restricted ensemble
(under our definition of discrepancy),
the integral is unbiased, {\it i.e.\/} $\avg{\eta}=0$, and
\bqa
N\avg{\eta^2} & = & 
\intk dx\left(f(x)\right)^2 - \left(\intk dx f(x)\right)^2\nl
& &
-\left(1-{1\over N}\right)\intk dx_1dx_2\;f(x_1)f(x_2)F_2(s;x_1,x_2)\quad,
\eqa
so that we may expect an improved error if $F_2$ is positive whenever
$x_1$ and $x_2$ are close together. For truly random points, the term
with $F_2$ drops out, and we recover the standard \mc\ result.\\

The treatment in \cite{first} has been purely formal; if it is to be any
good, we are obliged to show how it works out in practice.
The present paper is the first step in that direction: we shall apply
the formalism to a few particular definitions of $D_N$ in the case
of one-dimensional point sets and integrals. Of course this is more
or less academic since one-dimensional integrals can, in general, be
approximated to extreme accuracy by other, non-stochastic methods. But,
by keeping the extension to more dimensions in mind, we may hope to
develop useful results and insights. The lay-out of this paper is
as follows. In section 2, we briefly rehash the relevant results from
\cite{first}. In section 3, we present results for a very simple, and
not altogether too useful, discrepancy. In section 4, we proceed to
a more realistic definition of discrepancy, which will lead to a surprising
connection with well-known results for the star-discrepancy. We conclude
with some preliminary remarks on the extension to the case of
more-dimensional point sets and integrals.

\section{The formalism}
Given a point set of $N$ points $x_1,x_2,\ldots,x_N$ in the
unit interval $[0,1)$, we define its discrepancy as
\bq
D_N = {1\over N}\suml_{n>0}\si_n^2\suml_{k,l=1}^Nu_n(x_k)u_n(x_l)\quad,
\eq
where $u_{2n}(x) = \sqrt{2}\cos(2\pi nx)$,
$u_{2n-1}(x)=\sqrt{2}\sin(2\pi nx)$, and the numbers $\si_n$ depend on
the class of integrands that our particular integrand is thought to be
a member of. In the present model of integration, a typical integrand
$f(x)$ is of the form
\bq
f(x) = v_0 + \suml_{n>0}v_n u_n(x)\quad,
\eq
where the coefficients $v_n$ are distributed independently and normally
around zero with standard deviation $\si_n$. 
We shall always assume {\em translational invariance\/},
that is, $\si_{2n-1}=\si_{2n}$. This implies that
$F_2(s;x_1,x_2)$ only depends on $x_1-x_2$, so that we may
as well write it as $F_2(s;x_1-x_2)$. Moreover, for the integrands to
be quadratically integrable, the sum $\sum_{n>0}\si_n^2$
must converge. We define
\bqa
G_0(z) & = & \exp\left(-{1\over2}\suml_{n>0}(1-2z\si_n^2)\right)\quad,\nl
\phi(z;x) & = & \suml_{n>0}{2z\si_n^2\over1-2z\si_n^2}\;u_n(x)u_n(0)\quad.
\label{defs1}
\eqa
Let now the value of $D_N$ equal $s$ for our point set.
Then, we may compute $H_0$ and $F_2$ to leading order
in $1/N$ as follows:
\bqa
H_0(s) & = & {1\over2\pi i}\intinfi dz\;e^{-zs}\;G_0(z)\quad,\nl
R_2(s;x) & = &  {1\over2\pi i}\intinfi dz\;e^{-zs}\phi(z;x)\;G_0(z)\quad,\nl
F_2(s;x) & = & -\frac{R_2(s;x)}{H_0(s)}\quad,
\label{defs2}
\eqa
where the integration contours run to the left of all singularities.
Hence, both $H_0$ and $R_2$ vanish trivially whenever $s<0$.
Owing to the translational invariance in our model, the function $F_1$
vanishes to all orders in $1/N$, making the integration unbiased.
Moreover, the average value of $D_N$ for truly random point sets
is
\bq
\avg{D_N} = \suml_{n>0}\si_n^2\quad.
\eq
Finally, from the definitions (\ref{defs1},\ref{defs2}), 
we may immediately infer that
\bq
F_2(s;0) = {2\over H_0(s)}{\partial\over\partial s}
\left(sH_0(s)\right)\quad,
\label{maxf2}
\eq
which may serve as a check on the computations, and -- since $F_2$
may be supposed to reach a maximum for $x_1=x_2$, at least for small $s$
-- can yield an approximation to a lower bound on $s$, simply by putting
$F_2(s;x,x)=N$.

\section{A simple model}
\begin{figure}[tbp]
\begin{center}
\unitlength 1pt
\begin{picture}(410,202)(10,0)
%
%
%
\put(0,0){\epsffile{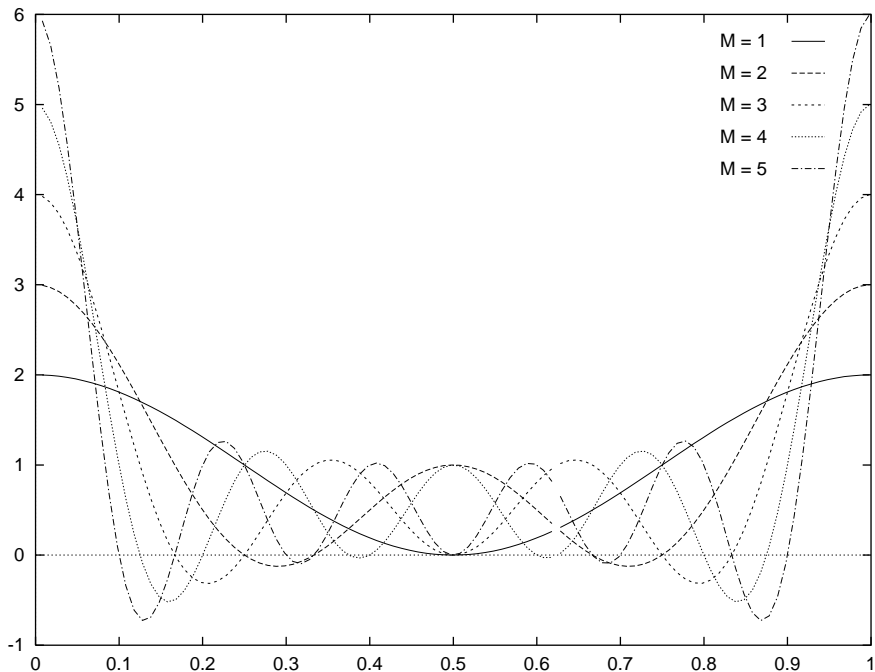}}
\end{picture}
\end{center}
\caption[]{The function $\xi_M(x)$ for various values of $M$.}
\label{fig:xim}
\end{figure}

The first, and simplest model, of integrands is obtained by taking
\bq
\si_{2n}^2 = \frac{1}{2M}\quad\;,\quad\;n=1,2,\ldots,M\quad,
\eq
and zero for higher values of $n$. This means that our typical
integrand has all Fourier components up to frequency $M$
occurring with equal strength on the average, higher modes
being absent. Note that this gives $\avg{s}=1$
for truly random point sets. In this case, we have
\bq
G_0(z) = \left(1-{z\over M}\right)^{-M}\quad\;,\quad\;
H_0(s) = {M^M\over\Gamma(M)}s^{M-1}e^{-sM}\quad,
\eq
where the last result (for $s>0$) follows when we close the integral
by a contour to the right. Note that, for large $M$, 
$H_0(s)$ approximates a Gaussian with unit mean and width $1/\surd{M}$.
The discrepancy is given by
\bqa
D_N & = & {1\over N}\suml_{k,l=1}^N\left[\xi_M(x_k-x_l)-1\right]\quad,\nl
\xi_M(x) & = & \suml_{n=0}^M\cos(2\pi nx)
= \frac{\cos(M\pi x)\sin((M+1)\pi x)}{\sin(\pi x)}\quad.
\eqa
A particularly interesting point set is the one with minimal
discrepancy, namely
\bq
x_k = \left\{x_0 + \frac{k}{N} \right\}\quad,\quad k=1,2,\ldots,N\quad,
\label{iets}
\eq
where $x_0$ is arbitrary, and $\{x\}$ denotes $x\;\mbox{mod}1$. The
discrepancy depends, in this case, only on $N$ and $M$:
\bq
D_N = 1 - {M \mbox{ mod} N\over M}\quad.
\eq
We see that, for $M<N$, the discrepancy actually vanishes. This
is reasonable, since an integrand that only has Fourier modes of 
frequency $N-1$ or lower is indeed integrated with zero error by
$N$ equidistant points. For large $M$, the value of $D_N$ lies
generally quite close to one.
The function $F_2(s;x)$ can easily be constructed in this case: 
since all the nonzero $\si_n$ are equal,it must be
proportional to $\xi_M(x)-1$, and from \eqn{maxf2} we can immediately find its normalization, to arrive at
\bq
F_2(s;x) = 2(1-s)\left[\xi_M(x)-1\right]\quad.
\eq
In Fig.~\ref{fig:xim}, we present $\xi_M(x)$ as a function of $x$ for
a few values of $M$. As $M$ increases, $\xi_M$ approaches $1/2+1/2\delta(\{x\})$, 
where $\delta(x)$ is the Dirac delta function.
Finally, the requirement $F_2(s;0)<N$ leads to
$s>1-N/(2M)$, which is a quite useless bound if $N\gg M$, but
is reasonable if $M\gg N$.\\

The main attraction of this model is its simplicity. 
The extension to more dimensions is straightforward.
Also, as proven in
the appendix to \cite{first}, one can indeed show that, relative
to the expected error from classical \mc, the expected error obtained
with a point set of discrepancy $s$ is indeed reduced by a factor $s$.
On the other hand, functions that have only a finite number of modes,
and all equally strong at that, do not really form a good model
for integrands likely to be encountered in practice. We therefore
move, now, to a more realistic model.

\section{A non-simple model}
The next model we shall study is characterized by
\bq
\si_{2n} = \frac{1}{n}\quad\;,\quad\;n=1,2,3,\ldots\quad.
\eq
Therefore, all Fourier modes can occur in the integrand, the higher
modes being typically smaller than the lowest ones. Note that, with
respect to the summability requirement on the $\si_n$, this is not
the slowest possible behaviour: we are, in principle, allowed to
take $\si_{2n}\sim n^{\epsilon-1/2}$, with any positive $\epsilon$:
but the above choice appears to be the most manageable. In fact,
integrands with discontinuities typically have Fourier coefficients
decreasing like $1/n$, so that the above choice for $\si$ appears to
be justified for integrands that may be discontinuous but in which
quadratically integrable singularities have been mapped away. This
is precisely the usual situation in, {\it e.g.}, particle phenomenology.

We may now reapply the formalism of section 2. In the first place,
the discrepancy is given by
\bqa
D_N & = & {1\over N}\suml_{k,l=1}^N\beta_1(x_k-x_l)\quad,\nl
\beta_1(x) & = & \suml_{n>0}\;{2\over n^2}\cos(2\pi nx)
\quad=\quad{\pi^2\over3}\left[1-6\{x\}(1-\{x\})\right]\quad,
\eqa
which follows trivially from the properties of Bernouilli
polynomials. Some special cases are of interest.
\begin{itemize}
\item The most non-uniform point set, in which all points coincide,
  has discrepancy $D_N = \pi^2N/3$.
\item For truly random points, the expected discrepancy is
  $\langle D_N\rangle = \pi^2/3$.
\item Generalized Richtmeyer sequences, defined by
  \bq
   x_k = \{ x_0 + k\alpha\}\quad,
  \eq
with arbitrary $x_0$ and non-integer $\alpha$, lead to a discrepancy
\bq
 D_N = {\pi^2\over3N}\left[N + 2\suml_{m=1}^{N-1}
 (N-m)(1-6\{m\alpha\}(1-\{m\alpha\}))\right]\quad,
\eq
which can conveniently be computed in time $\order{N}$.
\item The most uniform point set, that of \eqn{iets}, has a discrepancy 
\bq
D_N = {\pi^2\over3N}\quad.
\label{lowests}
\eq
\end{itemize}
The least and most uniform point sets are in fact, special cases of the
generalized Richtmeyer sequence with $\alpha=0$ and $\alpha=1/N$, respectively.
A nice illustration is the following. We consider generalized
Richtmeyer sequences, starting with $\alpha=0$. We then iterate
$\alpha\leftarrow 1/(1+\alpha)$, which gives us a series of rational
(Fibonacci) approximants to the golden ratio. For each $\alpha$ we
then compute the discrepancy, for several values of $N$. The results
are given in table 1. We see that the discrepancy decreases as the
approximation improves: however, for every fixed $N$ there are
some lower approximants that give a better discrepancy than some
higher approximants: the phenomenon of irregularity of distribution.
Similarly, with increasing $N$, the discrepancy decreases somewhat
slower than with $1/N$.
\begin{table}\begin{center}
\begin{tabular}{|l|l|r|r|r|}\hline\hline
iter. & $\alpha$ & $N=100$ & $N=1000$ & $N=10000$ \\ \hline
  1 &  1.00000000 &   328.98681 &  3289.86813 & 32898.68134 \\
  2 &   .50000000 &    82.24670 &   822.46703 &  8224.67033 \\
  3 &   .66666667 &    36.58333 &   365.54383 &  3655.40933 \\
  4 &   .60000000 &    13.15947 &   131.59473 &  1315.94725 \\
  5 &   .62500000 &     5.18977 &    51.40419 &   514.04190 \\
  6 &   .61538462 &     1.99806 &    19.46995 &   194.66713 \\
  7 &   .61904762 &      .78420 &     7.46330 &    74.60073 \\
  8 &   .61764706 &      .32284 &     2.85192 &    28.45961 \\
  9 &   .61818182 &      .16748 &     1.09343 &    10.87618 \\
 10 &   .61797753 &      .10234 &      .41917 &     4.15401 \\
 11 &   .61805556 &      .09481 &      .16257 &     1.58743 \\
 12 &   .61802575 &      .09296 &      .06628 &      .60668 \\
 13 &   .61803714 &      .09298 &      .02975 &      .23233 \\
 14 &   .61803279 &      .09287 &      .01554 &      .08931 \\
 15 &   .61803445 &      .09290 &      .00732 &      .03473 \\
 16 &   .61803381 &      .09288 &      .00832 &      .01404 \\
 17 &   .61803406 &      .09289 &      .00763 &      .00604 \\
 18 &   .61803396 &      .09289 &      .00784 &      .00303 \\
 19 &   .61803400 &      .09289 &      .00775 &      .00206 \\
 20 &   .61803399 &      .09289 &      .00779 &      .00146 \\
 21 &   .61803399 &      .09289 &      .00777 &      .00155 \\
 22 &   .61803399 &      .09289 &      .00778 &      .00149 \\
 23 &   .61803399 &      .09289 &      .00778 &      .00151 \\
 24 &   .61803399 &      .09289 &      .00778 &      .00150 \\
 25 &   .61803399 &      .09289 &      .00778 &      .00151 \\
\hline \hline
\end{tabular}
\caption[.]{Discrepancies for generalized Richtmeyer point sets with
$N=10^2,10^3,10^4$ points, with for $\alpha$ successive Fibonacci
approximations to the golden ratio.}
\end{center}\end{table}

\begin{figure}[tbp]
\begin{center}
\unitlength 1pt
\begin{picture}(410,202)(10,0)
%
%
%
\put(0,0){\epsffile{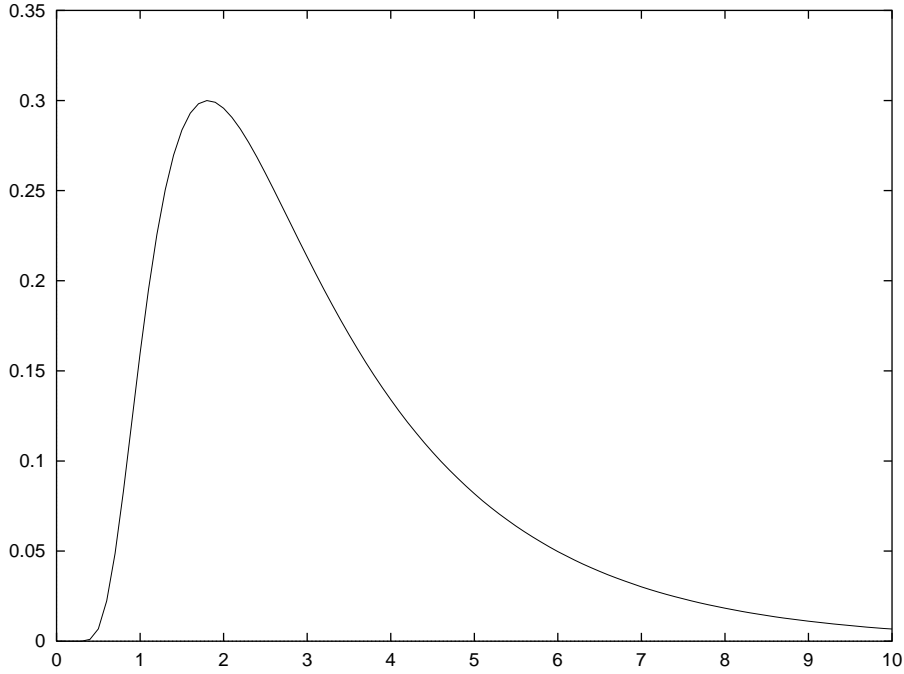}}
\end{picture}
\end{center}
\caption[]{The function $H_0(s)$ as a function of $s$.}
\label{fig:h0}
\end{figure}

The evaluation of $H_0$ and $F_2$ is somewhat more complicated than
in the previous model. In the first place, we have
\bq
G_0(z) = \prol_{n>0}{n^2\over n^2-2z} =
{\sqrt{2\pi^2z}\over\sin\sqrt{2\pi^2z}}\quad.
\eq
The simplest way to compute $H_0(s)$ is, then, to close the complex
contour of the $z$ integral to the right and contract it to run around
all the poles at $z=m^2/2$, $m=1,2,\ldots$: this immediately
gives us
\bq
H_0(s) = \suml_{m>0}(-)^{m-1}m^2e^{-sm^2/2}\quad.
\label{h0larges}
\eq
This result is, in some sense, surprising, since it is 
essentially identical to that of the star-discrepancy $D_N^\ast$:
by comparison with the Kolmogorov-Smirnov law, we have
\bq
\mbox{Prob}\left(D_N < s\right) =
\mbox{Prob}\left(D_N^\ast < \sqrt{s/(16N)}\right)\quad.
\eq
It is not completely clear why there should be such a simple
relation between the two cases, as they are based on quite
different models of the underlying class of typical integrands.
The result of \eqn{h0larges} is useful for large $s$ values, but
for small values another approach is more efficient. In it, we
write $z=-u^2/2$, with $u=a+iv$, where $a$ is fixed and $v$
runs over the real axis. This gives us
\bqa
G_0(z) & = & {i\pi u\over\sin i\pi u} =
2\pi u \suml_{m>0}e^{-(2m-1)\pi u}\quad,\nl
H_0(s) & = & \suml_{m>0}H_0(2m-1;s)\quad,\nl
H_0(y;s) & = & \intinf dv\;(a^2-v^2+2iav)
e^{\frac{1}{2}s(a+iv)^2-\pi y(a+iv)}\quad.
\eqa
We can now choose $a=y\pi/s$, independently for each $m$, and
so end up with
\bq
H_0(s) = \suml_{m>0}{\sqrt{2\pi}\over s^{5/2}}
\left(\pi^2(2m-1)^2-s\right)
e^{-{\pi^2(2m-1)^2\over2s}}\quad,
\label{h0smalls}
\eq
which converges very rapidly for small $s$. This is of course
the essence of the Poisson summation formula.
In Fig.~\ref{fig:h0}  we plot $H_0(s)$ as a function of $s$. The distribution
is seen to be considerably skewed, with the maximum
around $s\sim1.8$ occuring a good deal before $\avg{s}\sim3.28$.

\begin{figure}[tbp]
\begin{center}
\begin{picture}(410,202)(10,0)
%
%
%
%
  \put(347,0){\makebox(0,0)[lb]{$1$}}
\put(0,0){\epsffile{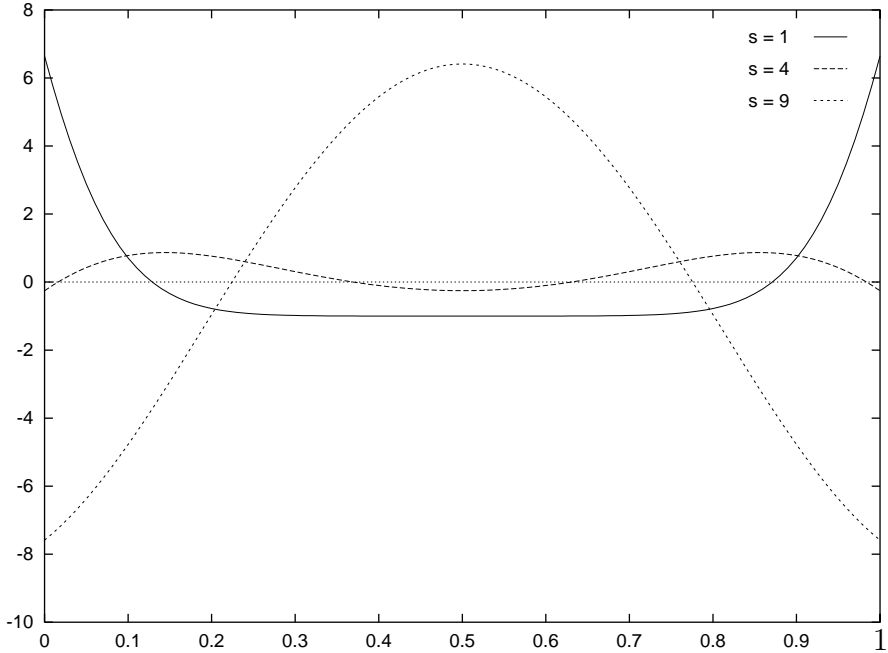}}
\end{picture}
\end{center}
\caption[]{The function $F_2(s;x)$ for various $s$.}
\label{fig:f2}
\end{figure}

\newpage
The computation of $R_2(s;x)$ runs along the same lines.
We may use the properties of Bernoulli polynomials to write
\bq
\phi(z;x) = \suml_{n>0}{4z\over n^2-2z}\cos(2\pi nx)
= 1 - G_0(z)\cos\left(\xi\sqrt{2\pi^2z}\right)\quad,
\eq
with $\xi=1-2x$. Therefore, we may write
\bqa
R_2(s;x) & = & H_0(s) - T(s;\xi)\quad,\nl
T(s;\xi) & = & {1\over2\pi i}\intinfi dz\;e^{-zs}
\cos\left(\xi\sqrt{2\pi^2z}\right){2\pi^2z\over\left(\sin(\sqrt{2\pi^2z})\right)^2}\quad;
\eqa
by closing the contour around the (double) poles, we obtain after
some delicate but straightforward algebra:
\bqa
R_2(s;\xi) & = & \suml_{m>0}(-)^{m-1}m^2e^{-sm^2/2}
\left[1 + (m^2s-3)\cos(2\pi mx)\right.\nl
& & \hphantom{\suml_{m>0}(-)^{m-1}m^2e^{-sm^2/2}}
\left. + \pi m(2x-1)\sin(2\pi mx)\right]\quad.
\label{r2larges}
\eqa
It is easily checked that $R_2$ vanishes upon integration over either
$x$ or $s$, as it should.
The form for small $s$ can be obtained along the same lines as above,
where we may use
\bq
\phi(-u^2/2;x)G_0(z) = G_0(z)
-2\pi^2u^2\left(e^{\pi u\xi}+e^{-\pi u\xi}\right)
\suml_{m>0}me^{-2\pi um}\quad,
\label{saddle}
\eq
to arrive at
\bqa
R_2(s;x) & = & H_0(s) - \suml_{m>0}m\,\left[T(2m+\xi)+T(2m-\xi)\right]\quad,\nl
T(y) & = & {\sqrt{2\pi}\over s^{7/2}}\,
\pi^2y\,(\pi^2y^2-3s)
e^{-{\pi^2y^2\over2s}}\quad.
\eqa
Again, also this form can, in principle, be obtained from 
\eqn{r2larges} by means of the Poisson summation formula. In Fig.~\ref{fig:f2}
we show the resulting forms for $F_2(s;x)$ for a number of different
values for $s$. We see that, for small $s$, this quantity indeed
peaks at $\{x\}\sim0$. For large $s$, on the other hand, it is
actually negative for small $x$, implying that an anomalously large
discrepancy occurs when the points are positively correlated, that
is, clumped together. Finally, for small $s$, it is easily
shown that $F_2(s;0)\sim\pi^2/s$. This gives us an
approximate lower bound of $\pi^2/N$ on $s$,
which should be compared to \eqn{lowests}. The bound is
quite reasonable, given that all our results are obtained in the
large-$N$ limit, where we have assumed $sN\gg1$. Including more
terms ought to enable us to improve on this limit.

Concerning the low-$s$ approximation, we want to remark the following.
We can, in principle, also obtain a similar result by a straightforward
saddle-point approximation, in which the integration over $z$ runs parallel
to the imaginary axis, with a large negative real part. That this approach
must work well is already evident from \eqn{saddle}, in which the leading term
is supplemented by exponentially small corrections. For large $s$, a similar
saddle-point approximation, where the real part of $z$ is close to $1/2$, is 
not so good, since the sub-leading corrections are of the same order as the 
leading term: this yields the correct behaviour with $s$, but with an erreoneous
normalization. However, we are mainly interested in point-sets with low $s$,
so that we may hope to at least approximate $H_0(s)$ and $F_2(s;x)$ by similar
saddle-point approximations~\cite{drie}.
This however will depend sensitively on the multi-dimensional generalization
of $\sigma_n$.

\newpage
\section{Conclusions}
In this paper, we have presented, for the first time, expressions for the 
two-point correlation function $F_2(s;x_1-x_2)$ for point sets, that are 
restricted to have a certain value $s$ of the discrepancy. These results
will be employed to study the improvement in integration error for 
Quasi-Monte Carlo~\cite{drie}.
A number of points remains to be more fully understood. 

On the one hand,
there is the similarity between our form of $H_0(s)$ and the Kolmogorov-Smirnov 
distribution. This is surprising, since they are based on quite different 
discrepancies. 

On the other hand,
our 'simple' model lends itself to a generalization to more dimensions, but
lacks realism. It remains to be seen whether there exists a discrepancy-definition,
that combines simplicity of results with an easy extension to higher dimensions.


\begin{thebibliography}{999}
\bibitem{first} J.~Hoogland and R.~Kleiss,
 {\it Discrepancy-based error estimates for \qmc. I: General formalism},
 preprint NIKHEF 96-02, e-Print archive: http://xxx.lanl.gov/hep-ph/9601270
\bibitem{drie} J.~Hoogland and R.~Kleiss, {\it in preparation}.
\end{thebibliography}
\end{document}